\documentclass[traditabstract]{aa} 
\usepackage{graphicx}
\usepackage{natbib}
\usepackage{txfonts}
\begin{document}
\title{The high-velocity cloud complex Galactic center negative as seen by EBHIS and GASS}
\subtitle{I. Cloud catalog and global properties}
\author{B. Winkel\inst{1,2}
          \and
         N. Ben Bekhti\inst{1}
           \and
         V. Darmst\"{a}dter\inst{1}
            \and
         L. Fl\"{o}er\inst{1}
            \and
         J. Kerp\inst{1}
             \and
         P. Richter\inst{3}
       }

\institute{Argelander-Institut f\"{u}r Astronomie (AIfA), Universit\"{a}t Bonn,
              Auf dem H\"{u}gel\,71, 53121 Bonn, Germany
	\and 
	    Max-Planck-Institut f\"{u}r Radioastronomie (MPIfR), 
              Auf dem H\"{u}gel\,69, 53121 Bonn, Germany
    \and
		Institut f\"{u}r Physik und Astronomie, Universit\"{a}t Potsdam, Karl-Liebknecht-Str.\,24/25, 14476 Golm, Germany\\
              \email{bwinkel@mpifr.de}
}

\date{??; ??}

\abstract{Using Milky Way data of the new Effelsberg--Bonn \ion{H}{i} Survey (EBHIS) and the Galactic All-Sky Survey (GASS), we present a revised picture of the high-velocity cloud (HVC) complex Galactic center negative (GCN). Owing to the higher angular resolution of these surveys compared to previous studies (e.g., the Leiden Dwingeloo Survey), we resolve complex GCN into lots of individual tiny clumps, that mostly have relatively broad line widths of more than $15\,\mathrm{km\,s}^{-1}$. We do not detect a diffuse extended counterpart, which is unusual for an HVC complex. In total 243 clumps were identified and parameterized which allows us to statistically analyze the data.  Cold-line components (i.e., $\Delta v_\mathrm{fwhm}<7.5\,\mathrm{km\,s}^{-1}$) are found in about 5\% only of the identified cloudlets. Our analysis reveals that complex GCN is likely built up of several subpopulations that do not share a common origin. Furthermore, complex GCN might be a prime example for warm-gas accretion onto the Milky Way, where neutral \ion{H}{i} clouds are not stable against interaction with the Milky Way gas halo and become ionized prior to accretion.}


\maketitle

\section{Introduction}
In recent years it has become more and more clear that the gas in the Milky Way (MW) halo plays an important role for understanding the evolution of our host galaxy. Matter is constantly expelled from the Galactic disk via winds and supernovae-driven outflows and at the same time material is accreted to the disk. Furthermore, gas produced by tidal interaction or remnants from the merging history of the MW may still reside in the halo \citep[e.g.,][ and references therein]{sancisi08}. Today, it is known that the so-called intermediate- and high-velocity clouds and complexes represent various stages of these processes \citep[e.g.,][ and references therein]{kalberla09}. A prime example for tidally disrupted material is the Magellanic Stream (MS) that extends over hundreds of degrees over the sky. In many other cases the origin of the gas is still debated.

One of these cases is the high-velocity cloud (HVC) complex Galactic center negative (GCN) which is one of the smaller cloud complexes \citep[in total flux and spatial size;][]{wakker91}. Owing to the lack of suitable background sources (stars or QSOs), relatively little is known about its chemical composition, gas properties, and distance. 

There are several early studies of complex GCN. The first \ion{H}{i} detection of an individual cloud belonging to GCN was described in \citet{saraber74}. This object is located close to the Galactic center ($(l,b)=(8\degr,-4\degr)$). Several origin scenarios were discussed in Saraber \& Shane's work, (1) an extra-galactic origin (with the cloud being a galaxy), (2) that it is an object penetrating the MW disk, and (3) that it is ejected from the Galactic center. In any case, the cloud would be located at a distance of several kpc (using some simplifications). The same cloud was independently studied by \citet{mirabel76}. They arrive at the conclusion that the object was ejected from the Galactic center and exclude other potential scenarios, e.g., that it is an extension of the Magellanic Stream, which is clearly contradicting \citet{giovanelli81}, who favored the MS scenario. Another GCN cloud was observed by \citet{cohen78}. Again, several origins were discussed, with an additional scenario coming into play, i.e., that the cloud is a local group object (but not a galaxy). 
\begin{figure*}
\centering
\includegraphics[width=0.98\textwidth,bb=14 40 691 532,clip=]{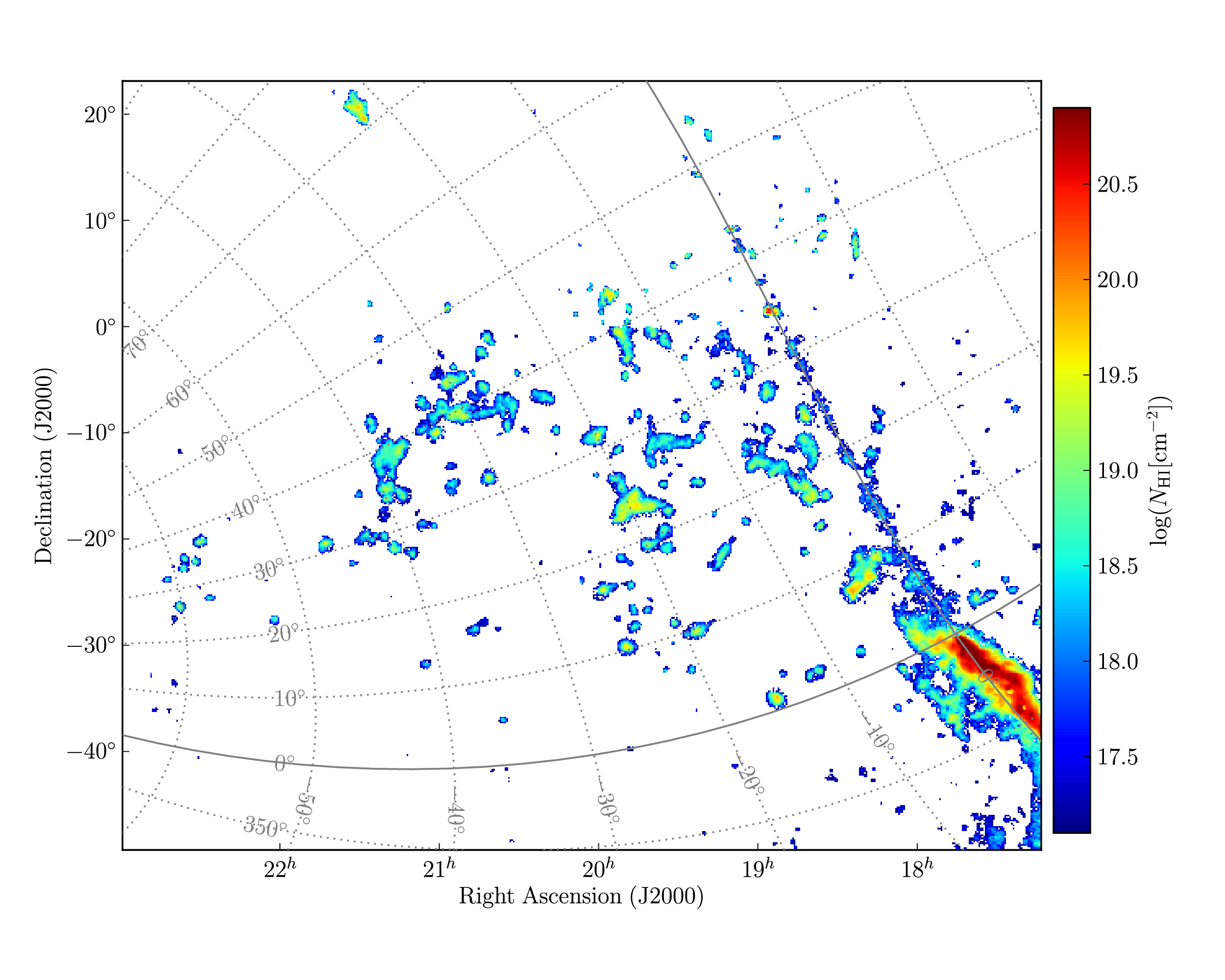}
\caption{Column density map of HVC complex GCN. Spurious emission is visible along the MW plane that is caused by the increased noise level in this direction. The column densities were computed over the velocity interval $v_\mathrm{lsr}=-360\ldots-160\,\mathrm{km\,s}^{-1}$ using a mask to suppress image artifacts caused by noise. The mask was created using a hanning filtered data cube (hanning-9) and applying a $2.5\sigma$ threshold. The graticule shows Galactic coordinates.}
\label{figclumpcolumndensitymap}%
\end{figure*}

Even the most recent analyses of the neutral hydrogen in GCN are based on observations dating back to about three decades ago \citep{bajaja85,hulsbosch88} which were performed with the Instituto Argentino de Radioastronomia 30-m and the Dwingeloo radio telescope. Both surveys were not fully sampled and hence are missing information. Furthermore, the small dish sizes led to a poor angular resolution of about $30\arcmin$. In these data sets GCN appears to be a population of widely scattered, small clouds at high negative velocities with sizes between 1 and 15 deg$^2$ \citep{wakker91}.  Most authors place GCN at a distance of more than about 10\,kpc such that the size of the complex is at least on the same order \citep[e.g.,][]{saraber74,wakker97,weiner01,wakker04} though the uncertainties are high because no direct distance estimates are available.

Using the X-ray shadowing technique, \citet{kerp99} could show that in the line of sight toward the galaxy Mrk\,509 (which is located in the GCN footprint) there is excess soft X-ray emission. For the same sight line \citet{sembach95} performed HST absorption-line measurements and detected highly ionized high-velocity gas. The X-ray data suggest that there is collisionally ionized gas, while \citet{sembach95} may have detected the cooler portion of the collisionally ionized plasma.

A few additional sight lines revealed highly ionized gas  \citep[e.g., \ion{C}{iv} and \ion{Si}{iv},][and references therein]{wakker01}. The presence of these metals provides strong evidence that GCN is not of primordial origin. Some authors conclude that GCN represents an inflow of gas to the Galactic center \citep{mirabel82,mirabel84} as well as the complex Galactic center positive (GCP), containing the famous Smith Cloud \citep[see][for a recent study]{lockman08}. Based on Leiden/Argentine/Bonn survey data \citep[LAB;][]{kalberla05}, \citet{jin10} used several clouds in the direction of complex GCN and calculated a hypothetical orbit that these objects might follow concluding that GCN likely was tidally stripped of a dwarf galaxy.

In this work we present new results obtained using the recently completed Galactic All-Sky Survey \citep[GASS;][]{mcclure2009,kalberla2010} obtained with the Parkes telescope and data from the ongoing Effelsberg--Bonn \ion{H}{i} Survey \citep[EBHIS;][]{winkel10a,kerp11} measured with the 100-m telescope at Effelsberg. The better angular resolution compared to previous large-area \ion{H}{i} surveys  and the full sampling of the data allowed us to obtain a revised picture of complex GCN.

The paper is organized as follows. In Section\,\ref{secdata} we describe the data from EBHIS and GASS that were used for the analysis. We compiled a catalog of clouds (Section\,\ref{seccatalog}) to perform a statistical analysis, the results of which are presented in Section\,\ref{secstatistics}. In Section\,\ref{secdiscussion} we discuss the results and conclude with a summary and outlook (Section\,\ref{secsummary}).

\section{Data}\label{secdata}
For our analysis we used data from the EBHIS and GASS surveys. In Fig.\,\ref{figclumpcolumndensitymap} we present a column density map covering the region $17^\mathrm{h}<\alpha<23^\mathrm{h}$, $-50\degr<\delta<25\degr$. This is the area in which we found \ion{H}{i} gas likely belonging to complex GCN. Above $\delta>-5\degr$ Fig.\,\ref{figclumpcolumndensitymap} shows EBHIS data, the remaining part is GASS data. We also checked the overlap regime between both surveys ($-5\degr<\delta<1\degr$) for consistency of the intensity calibration. Unfortunately, above  $\delta>25\degr$ EBHIS data were not yet available.
GASS data were reduced as described in \citet{kalberla2010}. The resulting angular resolution is $15\farcm6$, yielding an RMS of $57\,\mathrm{mK}$ per spectral resolution element ($\Delta v=0.8\,\mathrm{km\,s}^{-1}$, effective velocity resolution: $\delta v=1\,\mathrm{km\,s}^{-1}$). For EBHIS the data reduction scheme presented in \citet{winkel10a} was used. EBHIS has a higher nominal noise level of $\lesssim90\,\mathrm{mK}$ and lower spectral resolution (channel separation: $\Delta v=1.3\,\mathrm{km\,s}^{-1}$, effective velocity resolution: $\delta v=2.1\,\mathrm{km\,s}^{-1}$) but owing to the better angular resolution of $10\farcm5$, the resulting column density detection limit (after angular smoothing to the Parkes beam) is only about $\sqrt{2}$ higher\footnote{EBHIS will have similar sensitivity as GASS after the completion of the second coverage.}: $N^\mathrm{limit}_\ion{H}{i}=4.1\cdot10^{18}\,\mathrm{cm}^{-2}$ (GASS) and  $N^\mathrm{limit}_\ion{H}{i}=5.9\cdot10^{18}\,\mathrm{cm}^{-2}$ (EBHIS, unsmoothed: $8.9\cdot10^{18}\,\mathrm{cm}^{-2}$) calculated for a Gaussian-like emission line of width $20\,\mathrm{km\,s}^{-1}$ (FWHM) and a detection threshold of $5\sigma$. However, one has to keep in mind that (spatially) unresolved sources would be affected by beam-smearing degrading the detection probability, which in this respect favors the EBHIS survey because it has a better angular resolution. Both data sets were corrected for stray radiation using the method of \citet{kalberla80}, which was previously applied to the LAB survey.

\begin{figure}
\centering
\includegraphics[width=0.48\textwidth,clip=]{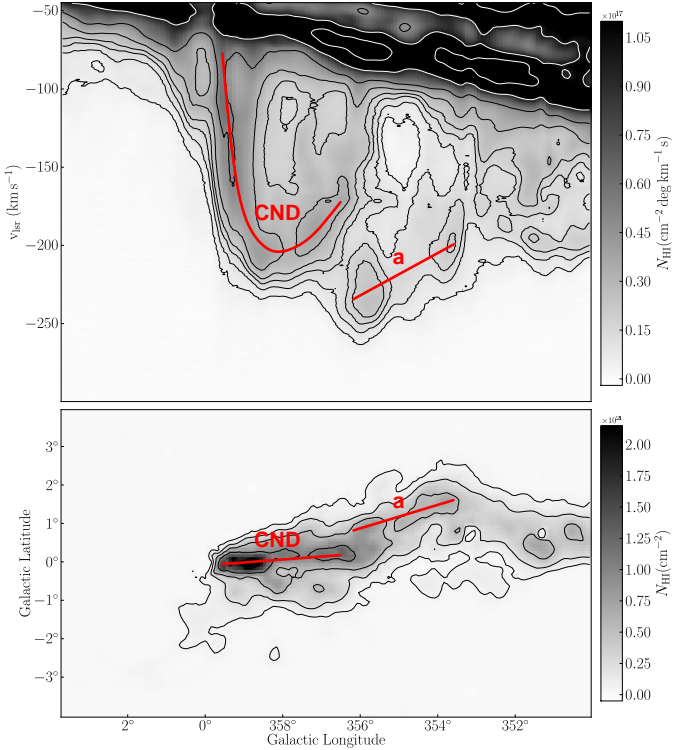}
\caption{Column density map of the negative velocity part of the central nuclear disk (CND) of the Milky Way (integrated over $v_\mathrm{lsr}=-290\ldots-160\,\mathrm{km\,s}^{-1}$) and an integrated position--velocity map ($-2\degr\leq b\leq+2\degr$). The structure labeled with (a) marks a large cloud association/filament located next to the CND (in position--velocity space). }
\label{figivcfeature}%
\end{figure}
 
Figure\,\ref{figclumpcolumndensitymap} shows the column density map of the data integrated over the velocity interval $v_\mathrm{lsr}=-360\ldots-160\,\mathrm{km\,s}^{-1}$.  Note that for $b\sim0\degr$ there is spurious emission visible because the spectra covering the Milky Way disk have a higher noise level and in some cases strong continuum sources may produce baseline artifacts. In the lower right of the column density map, at about $(l,b)=(-5\degr,0\degr)$, there is a bright feature that can be attributed to the central nuclear disk (CND) of the Milky Way. The CND is also prominently visible in CO data; see \citet{dame01}. A closer inspection of the data cube reveals an additional feature that cannot directly be attributed to the CND but may somehow be related to the GCN clouds. In Fig.\,\ref{figivcfeature} we show a column density map of the CND region (in Galactic coordinates, bottom panel) and a p--v diagram (top panel). The pronounced feature is marked with (a). We will come back to this in Section\,\ref{subsec:correlations}.

\section{Compilation of a cloud catalog}\label{seccatalog}
In order to build a catalog of clouds we searched for features in excess of $5\sigma_\mathrm{rms}$ in the original data cubes and hanning-filtered versions of the spectra. This spectral smoothing was performed with the miriad task \texttt{hanning} using  kernel widths of 3 and 9 spectral channels to optimize the search for cold and warm line components, respectively. However, we detected no additional features in the original and hanning-3 cube that were not apparent in the hanning-9 data cube, because all cold-line components that were found lie on top of a detection of a warm neutral cloud. For each identified clump we extracted the peak spectrum and used the GILDAS package \texttt{class} to fit Gaussians to the line profiles. If possible, two separate Gaussians were used to account for cold and warm (or multiple) components. 

Based on our cloud sample we find that complex GCN extends from $l\sim0\degr\ldots70\degr;~b\sim-60\degr\ldots10\degr$ \citep[and not as previously reported from $l\sim0\degr\ldots50\degr;~b\sim-40\degr\ldots10\degr$;][and references therein]{wakker91} and covers a fairly wide velocity range between $-350<v_\mathrm{lsr}\lesssim-70\,\mathrm{km\,s}^{-1}$. The latter limit is not very well defined; at velocities of $v_\mathrm{lsr}>-200\,\mathrm{km\,s}^{-1}$ the CND feature and its environment overlap with the GCN. Therefore, to classify whether a clump is part of GCN, we chose a hard limit of $v_\mathrm{lsr}<-70\,\mathrm{km\,s}^{-1}$ with the additional constraint that GCN clumps must be well-separated in position--velocity space from the CND. For the subsequent analysis we converted radial velocities to the Galactic standard of rest (GSR) frame, which is better suited for objects located in the MW halo, especially given the large  spatial extent of complex GCN. Note that \citet{wakker04} detected GCN clouds in a smaller LSR velocity range of $-350<v_\mathrm{lsr}\lesssim-170\,\mathrm{km\,s}^{-1}$. The newly identified clouds with $v_\mathrm{lsr}>-170\,\mathrm{km\,s}^{-1}$ lie mainly at high (negative) Galactic latitudes and have similar properties and more importantly are consistent with the GSR velocities of the previously known clumps.

In total we were able to identify and parameterize 243 clumps (48 EBHIS, 195 GASS). This is about twice the number of objects detected in the most recent survey of GCN using LAB data \citep{kalberla06}, which can be attributed mainly to the superior angular resolution and full spatial sampling, because most of the newly found clouds are still not resolved even with the smaller beam sizes of the Effelsberg and Parkes telescopes.

Many of the clouds are connected, forming larger structures or filaments in the column density map; compare Fig.\,\ref{figclumpcolumndensitymap}. However, the inspection of the data cubes shows that many of the components of these features are well-separated in velocity and the filaments are likely just the superposition of several more or less isolated clumps. It is important to keep in mind that the column density map shows the integrated intensity over a huge velocity range. Almost all clumps have a very small angular diameter of less than two beam widths. In a few cases we also identify head-tail- or  horse-shoe-like morphologies \citep{meyerdierks91,bruens00}, suggesting ram-pressure interaction. 

Interestingly, we find almost no extended diffuse gas in the 21-cm line emission surrounding the compact clumps. Because the column densities of a substantial fraction of the clumps are high both in terms of absolute value and signal-to-noise ratio, it is very unlikely that this is only a detection-limit effect. 
\begin{figure*}
\centering
\includegraphics[width=0.98\textwidth,bb=137 448 867 1166,clip=]{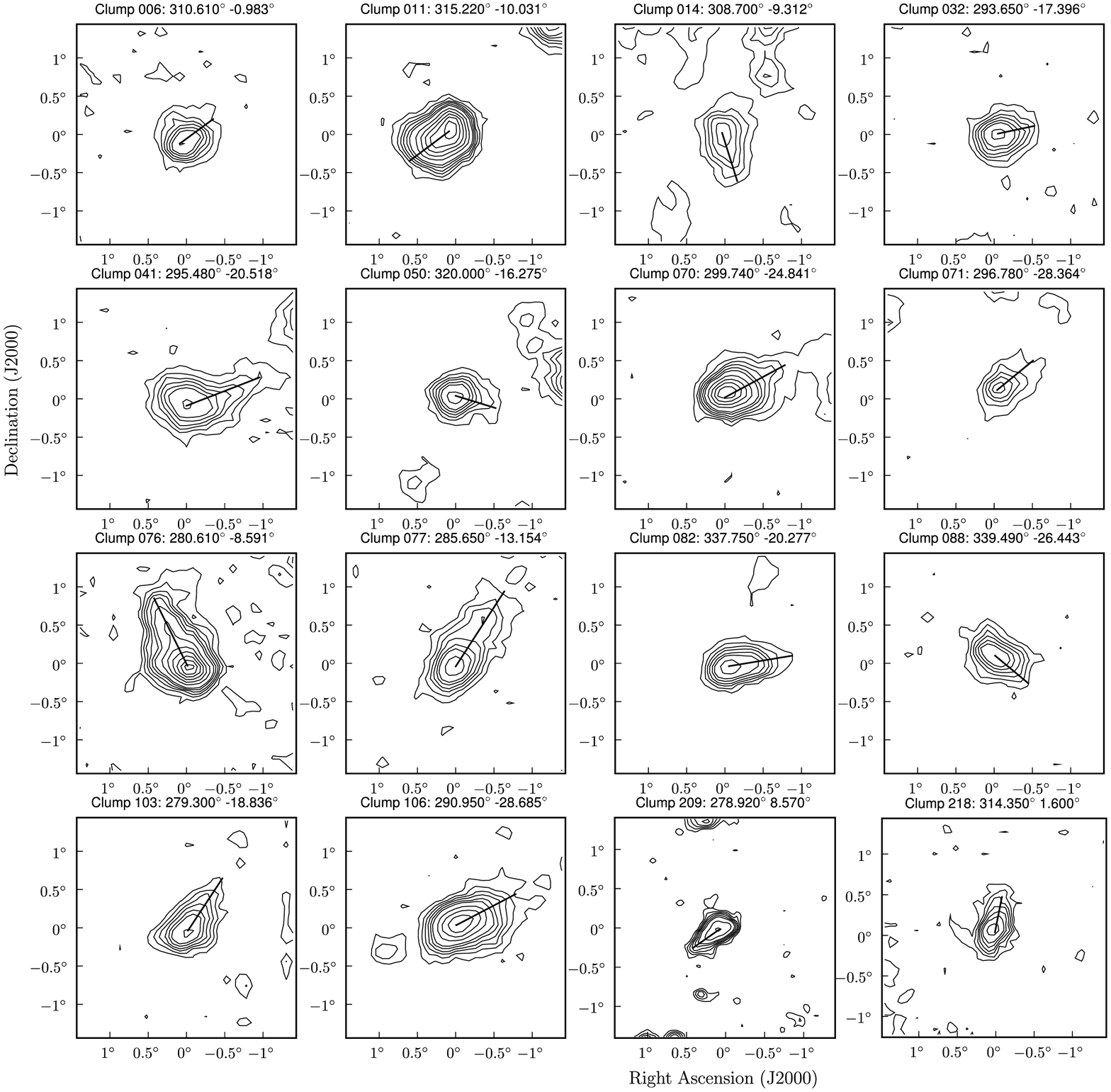}
\caption{In total 33 of the isolated clumps show head-tail like morphologies. The figure shows column density maps of 16 of them (Clump 209 and 218 are EBHIS detections, the remaining clouds shown are from GASS data). The remaining 17 cannot as clearly be identified as HT structures in the column density maps, because nearby clumps partly overlap in the projection. They were confirmed to likely be HT clouds by a close inspection of the data cube. Contours are $\left[3,6,9,12,16,20,30,40,\ldots \right]\cdot\sigma_\mathrm{rms}$ with $\sigma_\mathrm{rms}=5\cdot10^{17}\,\mathrm{cm}^{-2}$ (GASS) and $\sigma_\mathrm{rms}=7\cdot10^{17}\,\mathrm{cm}^{-2}$ (EBHIS). }
\label{figHTmomentmaps}%
\end{figure*}
Several of the clouds show signs of ongoing interaction but some of these are part of filaments or larger structures, which makes an analysis complicated. For 33 of the clouds we identified head-tail-like morphologies. In Fig.\,\ref{figHTmomentmaps} we show column density maps of the 16 head-tail (HT) clouds that are more or less isolated. The black solid lines mark the vector between the head and tail. For the remaining clumps the column density maps alone do not allow for an unambiguous definition of the HT vector. Therefore, we inspected the data cube for a better distinction of head, tail, and adjacent clumps, the latter of which are superposed on the HT cloud in the column density maps but are sufficiently separated in position--velocity space. The plot also clearly shows that even at the higher angular resolution of the EBHIS the clumps remain mainly unresolved. The HT clouds have to be treated with care because the superposition of two or more unresolved small clumps could easily mimic head-tail structures.

\section{Statistical analysis of the cloud sample}\label{secstatistics}
\subsection{General properties of GCN as a whole}
\begin{figure}
\centering
\includegraphics[width=0.45\textwidth,bb=66 56 400 191,clip=]{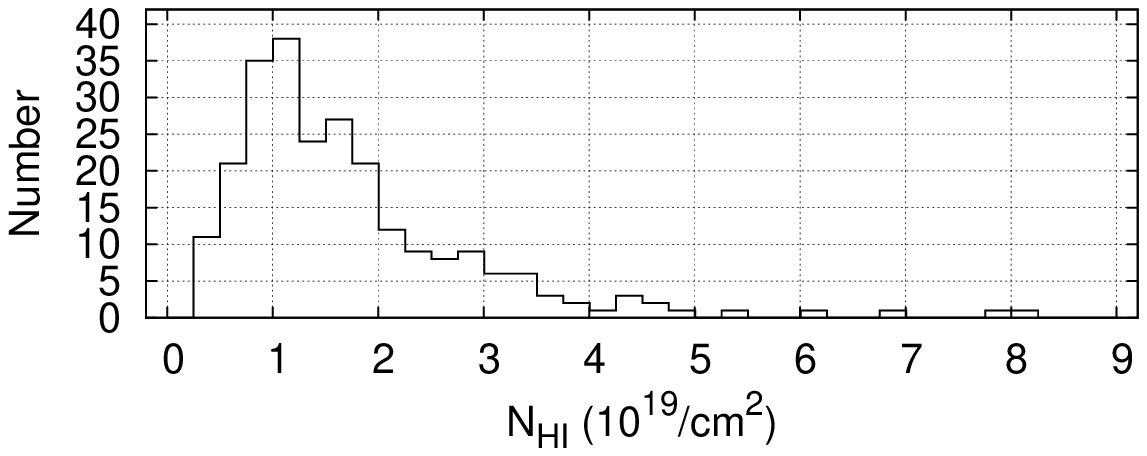}\\[1ex]
\includegraphics[width=0.45\textwidth,bb=66 56 400 191,clip=]{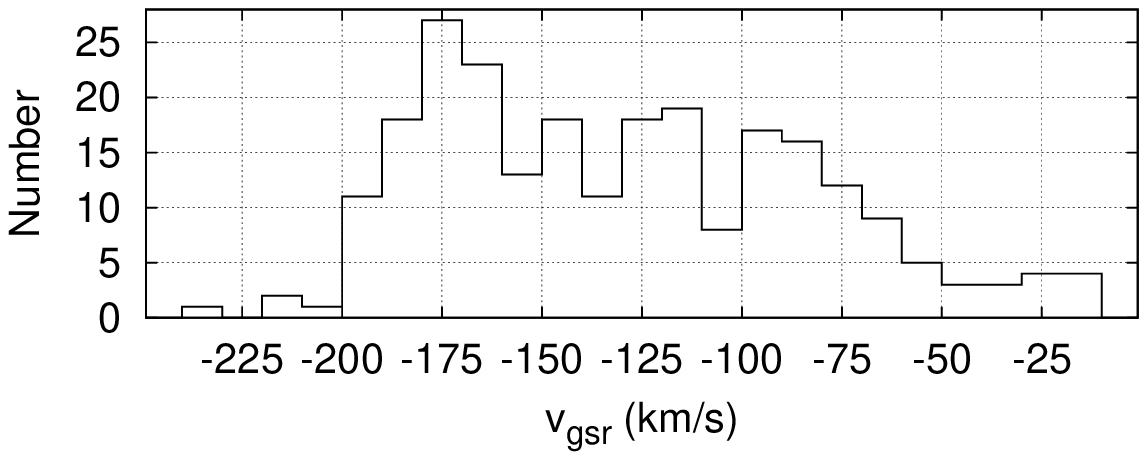}\\[1ex]
\includegraphics[width=0.45\textwidth,bb=66 56 400 191,clip=]{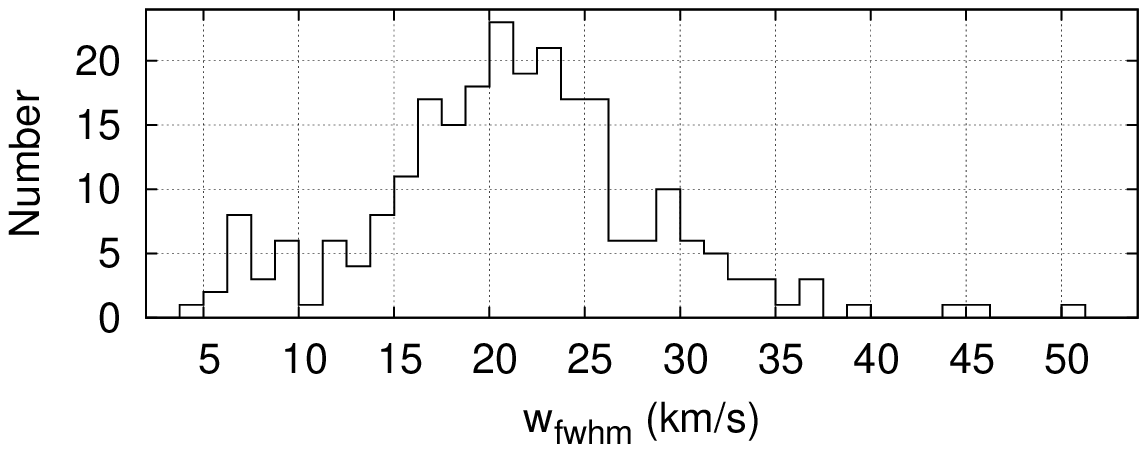}
\caption{Histogram of column densities, mean radial velocities, and  line widths (FWHM) of the cloud sample.}
\label{fighisto1}%
\end{figure}

In Fig.\,\ref{fighisto1} we show the distribution of the column densities, radial velocities, and line widths (FWHM). Column densities lie in the range between $3\cdot10^{18}\,\mathrm{cm}^{-2}$ and $8\cdot10^{19}\,\mathrm{cm}^{-2}$ with a median of $N_\ion{H}{i}=1.4\cdot10^{19}\,\mathrm{cm}^{-2}$. The line width histogram reveals that the bulk of the clumps has relatively broad profiles (median value is $\sim21\,\mathrm{km\,s}^{-1}$) and only few sources with a cold-gas component were found with $\Delta v<7.5\,\mathrm{km\,s}^{-1}$ in only 11 of 243 cases (i.e., less than 5\%). This is a very exceptional case compared to most other HVC complexes \citep[e.g.,][]{kalberla06}, especially considering the fact that we find many unresolved HVCs and no evidence for a diffuse \ion{H}{i} component in the 21-cm emission. While in principle the lack of cold cores might be a detection problem caused by beam-filling, many other HVC complexes --- even at farther distances --- do reveal a substantial fraction of cold cores. If the typical size of cold cores in GCN is not exceptionally small, which we consider unlikely, the lack of a cold component has to be real. Note that our finding agrees well with \citet{kalberla06}, who performed a statistical analysis of multicomponent structures in HVCs based on the LAB survey. Some of our sight lines show line widths that can hardly be explained by a pure Doppler broadening. For example $\Delta v_\mathrm{fwhm}\gtrsim30\,\mathrm{km\,s}^{-1}$ converts to $T_\mathrm{D}\gtrsim2\cdot10^5\,\mathrm{K}$ which is too high for a WNM cloud. This again hints at the presence of substructure not resolved with the telescope beams of Parkes and Effelsberg. 

Interestingly, other recent studies of the MW halo using high-resolution instruments report on numerous small clumps as well. \citet{stanimirovic08} detected nearly 200 of these objects at the tip of the Magellanic Stream in precursor observations of the Galactic Arecibo L-band Feed Array \ion{H}{i} survey \citep[GALFA-HI;][]{peek11}. However, the authors report on a (somewhat) higher fraction of cold-line components of 12\%. It will surely be interesting to re-analyze other HVC complexes once the new survey data are available.

\begin{figure}
\centering
\includegraphics[width=0.49\textwidth,bb=75 76 356 360,clip=]{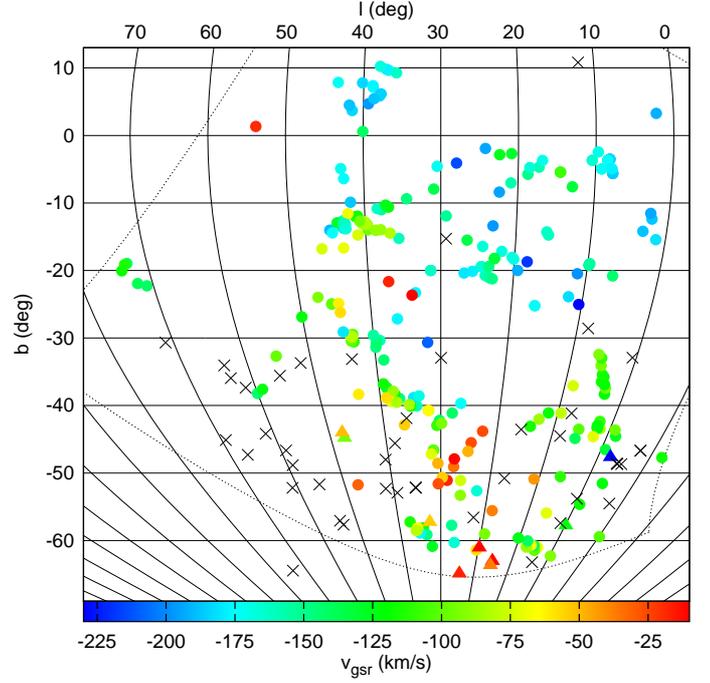}
\caption{Radial velocities and spatial positions of clouds in our catalog.  We additionally had access to the metal absorption line sample of \citet{benbekhti08}, who searched for \ion{Ca}{ii} and \ion{Na}{i} gas in QSO spectra obtained with UVES/VLT; see Section\,\ref{subsubsec:qso}. Nondetections are marked with black crosses, \ion{Ca}{ii} detections are marked with filled triangles (velocity is color-coded). No \ion{Na}{i} absorption lines were detected. The dotted line shows the area that we used for the analysis corresponding to the map in Fig.\,\ref{figclumpcolumndensitymap}.}
\label{figclumpcoordinates}%
\end{figure}

The distribution of clouds in the three-dimensional positional parameter space ($l,\,b,\,v_\mathrm{gsr}$) is extremely wide-spread. In Fig.\,\ref{figclumpcoordinates} we show the position of all detected clumps (in Galactic coordinates) with their radial velocity color-coded. Clouds in the northeastern part tend to have generally higher absolute values of radial velocity, whereas the lowest absolute values are found in the south. However, a close inspection of the data shows that even within strongly confined patches\footnote{For example at $(l,b)\sim(40\degr,-13\degr),\,(38\degr,-38\degr)$.} (of few degrees size) the scatter in radial velocities is large --- they are in many cases distributed over more than $100\,\mathrm{km\,s}^{-1}$.

\subsubsection{Column density distribution}
\begin{figure}
\centering
\includegraphics[width=0.4\textwidth,bb=58 60 395 291,clip=]{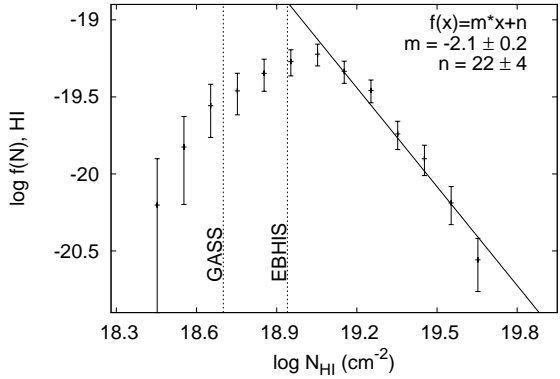}
\caption{Column density distribution function of the cloud sample using logarithmic binning. The vertical dotted lines mark the $5\sigma$ column density detection limit of the data (GASS: $N_\ion{H}{i}^\mathrm{limit}\approx5.2\cdot10^{18}\,\mathrm{cm}^{-2}$; EBHIS: $N_\ion{H}{i}^\mathrm{limit}\approx8.9\cdot10^{18}\,\mathrm{cm}^{-2}$ for Gaussian-like emission of $20\,\mathrm{km\,s}^{-1}$ line width), leading to an incompleteness of the sample. The regime $N_\ion{H}{i}\geq10^{19}\,\mathrm{cm}^{-2}$ can be described by a power law with a slope of $-2.1\pm0.2$.}
\label{figcdd}%
\end{figure}

\begin{figure*}
\centering
\includegraphics[width=0.4\textwidth,clip=]{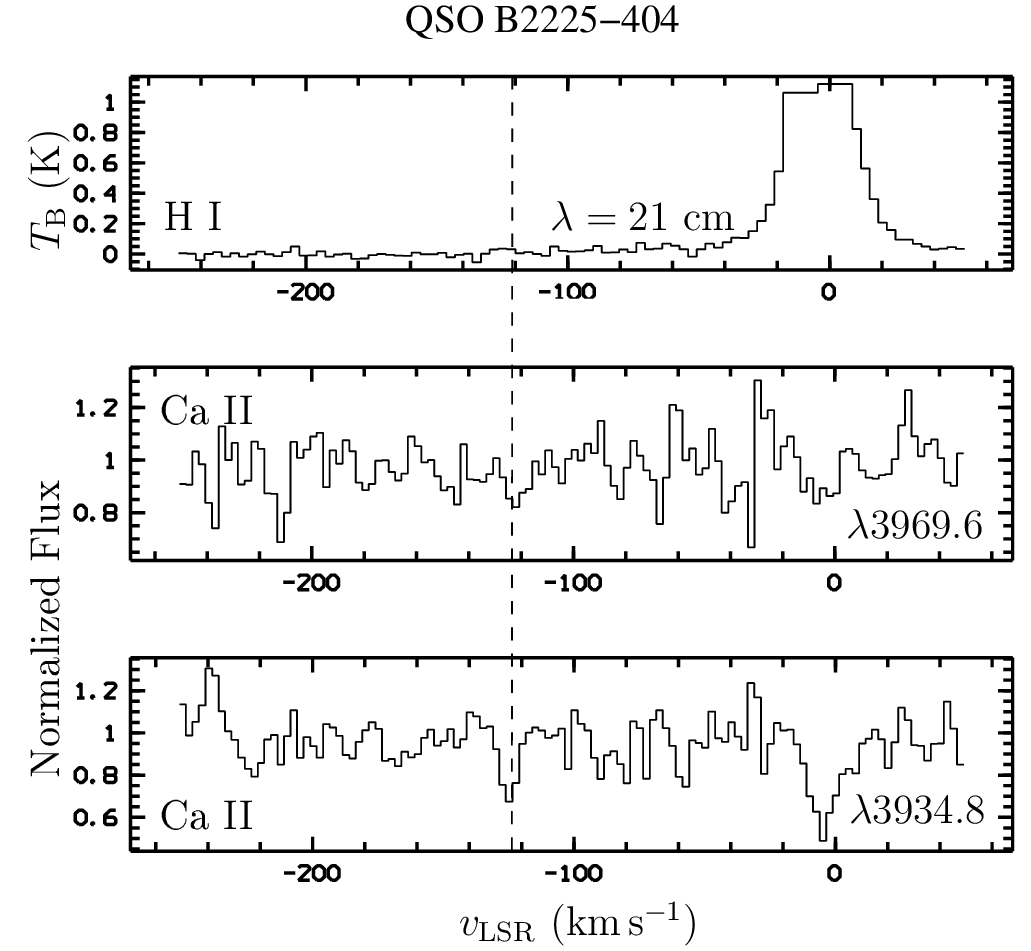}\qquad
\includegraphics[width=0.4\textwidth,clip=]{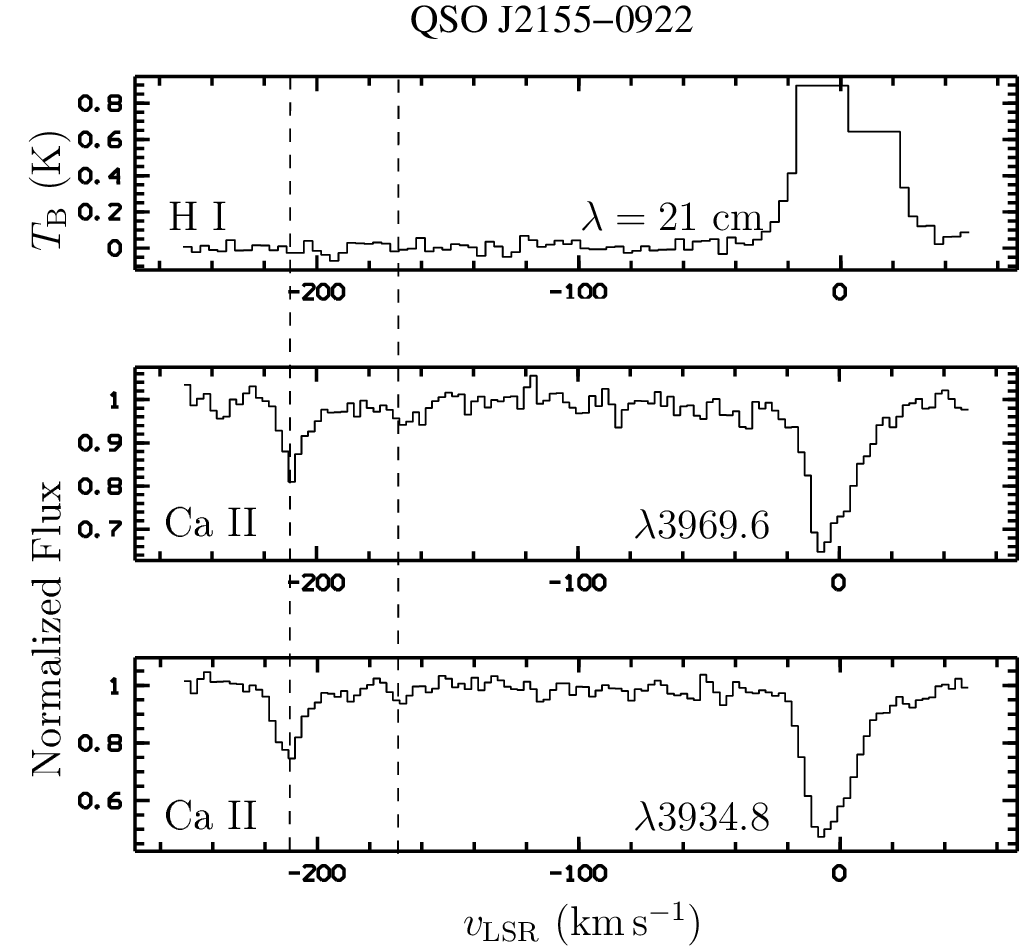}
\caption{\ion{Ca}{ii} absorption spectra in the direction of the quasars QSO\,B2225$-$404 and QSO\,J2155$-$0922. The upper panels show the corresponding \ion{H}{i} spectra that were spectrally binned to improve the signal-to-noise ratio. Note that Milky Way emission was clipped. In both cases no \ion{H}{i} detection was made. The dashed vertical lines mark the LSR velocities of the absorbers.}
\label{figcaIIspectra}%
\end{figure*}

\citet{churchill03} defined the column density distribution function (CDD), i.e., the number of detected clouds per column density interval\footnote{To obtain proper normalization, each value was divided by the total number of clouds in the sample and the bin width of each column density interval. The latter ensures independence of the power law slope on the chosen binning scheme (e.g., linear or logarithmic).}. Fig.\,\ref{figcdd} shows the CDD as derived from our sample. As expected, toward lower values of $N_\ion{H}{i}$ the catalog becomes increasingly incomplete. For the range $N_\ion{H}{i}\geq10^{19}\,\mathrm{cm}^{-2}$ a power law was fitted to the CDD, yielding a slope of $-2.1\pm0.2$. We will not compare this value to other studies at this point, because the beam size of the instrument is thought to have some impact on the results. Nevertheless, the obtained slope might serve as reference for future analyses of HVCs based on EBHIS and GASS.

Although it is tempting to interpret the slow drop-off above the nominal detection limits of the surveys as a physical effect, one must be very careful. As stated above, the column density detection limits were calculated for the very particular case of $20\,\mathrm{km\,s}^{-1}$ line width. For clouds with higher line widths the limit increases (proportional to $\sqrt{\Delta v_\mathrm{fwhm}}$). More importantly, the detection thresholds are very sensitive to the angular size of spatially unresolved clumps. For example, for an object with a size that matches the EBHIS beam, the GASS detection limit is enlarged by about $(15\farcm6/10\farcm5)^2\approx2.2$. From our sample we can safely assume that a high fraction of clouds is still below the angular resolution of both surveys. Consequently, the detection probability is seriously affected by beam filling problems. Based on our sample it is not possible, unfortunately, to conclude whether the drop-off is real or not.

\subsubsection{Metal absorption spectroscopy}\label{subsubsec:qso}

Important information on the physical properties of complex GCN comes from UV absorption-line measurements. We had access to the extensive metal absorption line sample of \citet{benbekhti08}, who searched for \ion{Ca}{ii} and \ion{Na}{i} gas in QSO spectra obtained with UVES/VLT. In total 57 QSO lines of sight were available for the region of interest, but unfortunately, they almost exclusively trace the southern part of complex GCN. From these 57 sight lines 10 show \ion{Ca}{ii} absorption with radial velocities that agree well with the \ion{H}{i} emission; compare Fig.\,\ref{figclumpcoordinates}. Only one feature at $(l,b)=(-2\degr,-48\degr)$ has a much higher absolute velocity compared to adjacent (\ion{H}{i}) clumps.  The median \ion{Ca}{ii} column density is about $N_\ion{H}{i}\sim5\cdot10^{11}\,\mathrm{cm}^{-2}$, the maximal value  was found at $(l,b,v_\mathrm{gsr})=(0\degr,-58\degr,-120\,\mathrm{km\,s}^{-1})$ and has $N_\ion{H}{i}\sim3\cdot10^{12}\,\mathrm{cm}^{-2}$. In Fig.\,\ref{figcaIIspectra} we show spectra for two of the sight lines. In both cases no \ion{H}{i} detection was made.

Remarkably, no \ion{Na}{i} absorption lines were detected. \ion{Na}{i} is a tracer for the colder dense cores of clouds, while \ion{Ca}{ii} is usually found in the warmer envelopes of clumps. For comparison, in the complete QSO absorption line sample in about 40\% of the cases \ion{Na}{i} was detected associated to \ion{Ca}{ii} absorbers. A second noteworthy point is that in most cases no \ion{H}{i} emission is seen directly on top of the absorption features. This suggests that there is a lot of material below the detection limit of the \ion{H}{i} surveys --- metal absorption spectroscopy is a very sensitive tool to detect the low-column density portion of the gas.

\begin{figure}
\centering
\includegraphics[scale=0.85,bb=147 514 305 710,clip=]{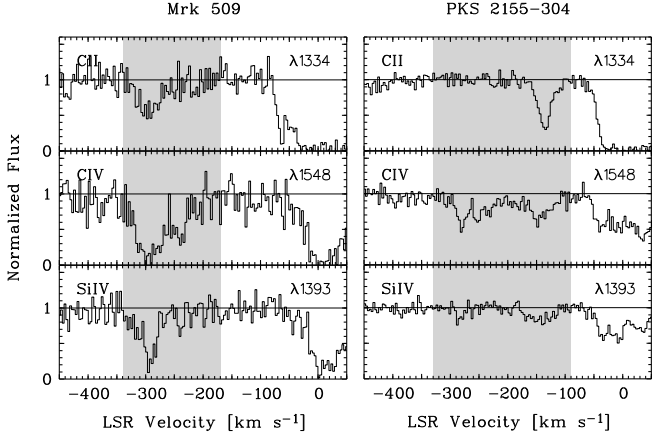}
\includegraphics[scale=0.85,bb=174 301.3 305 501.3,clip=]{fig8.eps}
\caption{Absorption profiles of \ion{C}{ii}, \ion{C}{iv}, and \ion{Si}{iv} toward Mrk\,509 (left panel) and PKS\,2155 (right panel), based on archival HST/STIS data.}
\label{figstisdata}%
\end{figure}

Using spectral data from the Goddard High Resolution Spectrograph (GHRS) on HST, \citet{sembach99} detected absorption by intermediate and high ions such as C\,{\sc ii}, Si\,{\sc iii}, C\,{\sc iv}, and Si\,{\sc iv} in the velocity range between $-200$ and $-300\,\mathrm{km\,s}^{-1}$ toward the quasar Mrk\,509; $(l,b)=(36.0\degr,-29.9\degr)$. We show in Fig.\,\ref{figstisdata} (left panel) the absorption profiles of \ion{C}{ii}, \ion{C}{iv}, and \ion{Si}{iv} toward Mrk\,509, based on archival HST/STIS data \citep[see also][]{richter09}. 

The Mrk\,509 sightline intersects complex GCN in a region that is close to a larger population of \ion{H}{i} clouds seen in 21-cm emission (see Fig.\,\ref{figclumpcolumndensitymap}). No 21-cm emission is seen toward Mrk\,509 itself, so that \citet{sembach99} classify this high-ion absorber toward Mrk\,509 as ``highly-ionized HVC''. From our 21-cm data it is now evident that the high-ion absorption toward Mrk\,509 traces the ionized component of complex GCN. From a photo-ionization model of the observed ion ratios toward Mrk\,509, \citet{sembach99} derive characteristic gas densities of $n_\mathrm{H}\sim 10^{-4}\,\mathrm{cm}^{-3}$, gas temperatures of $T=2\ldots3\cdot 10^4\,\mathrm{K}$, and thermal pressures of $P/k=1\ldots5\,\mathrm{cm}^{-3}\,\mathrm{K}$ for the ionized gas toward Mrk\,509. The total column density of ionized hydrogen in the HVC is estimated to be $\log N_\ion{H}{ii}=18.5\ldots19.7\,[\mathrm{cm}^{-2}]$, depending on the metallicity of the gas.

The second GCN sightline that has been studied in UV absorption by \citet{sembach99}
is the one toward PKS\,2155$-$304; $(l,b)=(17.7\degr,-52.2\degr)$. It passes through an outer region of complex GCN that is completely devoid of \ion{H}{i} 21-cm emission (see Fig.\,\ref{figclumpcolumndensitymap}). Toward PKS\,2155$-$304 high-velocity absorption is seen between $-90$ and $-330\,\mathrm{km\,s}^{-1}$; see Fig.\,\ref{figstisdata} (right panel). The intermediate- and high-ion absorption in complex GCN between $-170$ and $-330\,\mathrm{km\,s}^{-1}$ is much weaker toward PKS\,2155$-$304 compared to the Mrk\,509 sightline, suggesting that the ionized (and thus the total) gas mass is substantially lower in this direction. Another prominent absorption feature
that is possibly related to complex GCN is present near $-130\,\mathrm{km\,s}^{-1}$. Here, the weak intermediate-and high-ion absorption is accompanied by relatively strong absorption in \ion{Si}{ii} and \ion{C}{ii} \citep[see also][their Fig.\,3]{sembach99}.

Some interesting conclusions can be drawn from these numbers. For instance, the Mrk\,509 data suggest that the column density of ionized hydrogen in complex GCN is substantial (covering a much wider spectral range than traced by 21-cm emission), possibly even higher than the (typical) neutral gas column density in the \ion{H}{i} clumps detected with EBHIS and GASS. This is supported by the detection of high-velocity \ion{O}{vi} absorption toward Mrk 509, which extends from $-90$ to $-350\,\mathrm{km\,s}^{-1}$ \citep{wakker03}.

\subsection{Evidence for a tidal gas origin of GCN}

\subsubsection{Possible association with the Magellanic System}
\begin{figure}
\centering
\includegraphics[width=0.49\textwidth,bb=75 76 356 360,clip=]{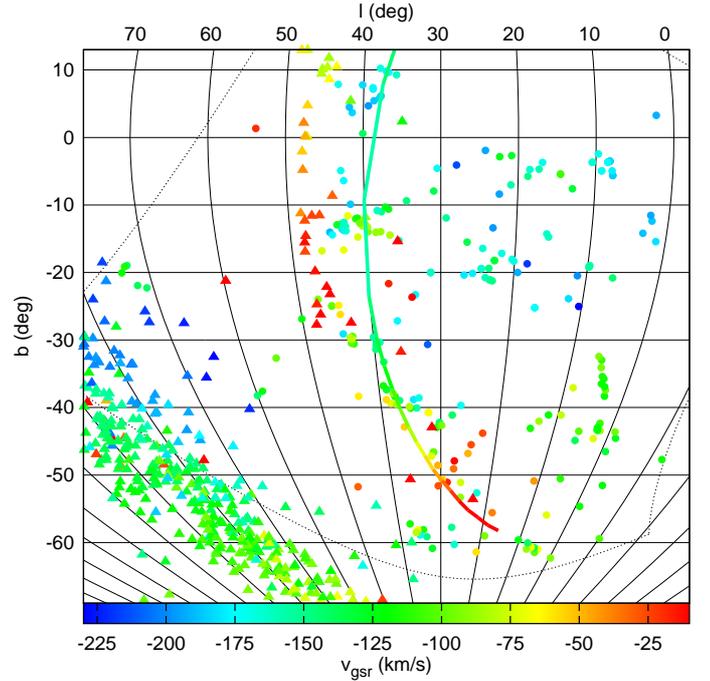}
\caption{Radial velocities and spatial positions of clouds in our catalog as in Fig.\,\ref{figclumpcoordinates} (filled circles). Diaz \& Bekki kindly allowed us access to their GN96+drag simulation data \citep[filled triangles;][]{diaz11}. The solid line is a proposed orbit for some of the GCN clouds based on LAB data as calculated by \citet{jin10}.}
\label{figclumpcoordinatesGN96}%
\end{figure}

N-Body simulations of the Magellanic System including drag forces \citep[GN96+drag model;][]{diaz11} predict gas in the region covered by GCN \citep[figure\,2 of][]{diaz11}. Because \citet{diaz11} only provide the distribution of GSR velocities and galacto-centric distances projected to (Magellanic) longitude in their paper, the authors kindly gave us access to their data, which allows us to compare the three-dimensional ($l$, $b$, $v_\mathrm{gsr}$) distribution of the simulated N-Body particles with our sample; see Fig.\,\ref{figclumpcoordinatesGN96}. 

Apparently the spatial position of the long ``filament'' (in the N--S stripe) almost matches the simulated data. However, the simulated velocities are offset by about $+100\,\mathrm{km\,s}^{-1}$ when compared to the median value of the data --- though the scatter along the filament is huge ($\sim100\,\mathrm{km\,s}^{-1}$ difference between minimal and maximal values). The large scatter itself is also not predicted, the simulated stream of gas has a scatter of about $\sim25\,\mathrm{km\,s}^{-1}$.

Furthermore, the two large cloud regions in the northeastern and southeastern parts are not predicted by the simulations. On the other hand the simulations predict a lot of gas (with $v_\mathrm{gsr}\lesssim-175\,\mathrm{km\,s}^{-1}$) in the southwestern part of the area, where we do not detect any cloud in the investigated area (marked with the dotted line in Fig.\,\ref{figclumpcoordinatesGN96}). We also briefly checked the GASS data for $\alpha>23^\mathrm{h}$ (beyond the southeastern edge of the observed area) and confirm the presence of gas belonging to the MS ($v_\mathrm{gsr}\sim-125\,\mathrm{km\,s}^{-1}$) as visible in the figure. Note that in the simulations the filament is an extension of the Leading Arm, while the gas in the southwest is associated to the Magellanic Stream. Consequently, the distances of both features differ. The former has distances between 14 and 22\,kpc (higher values in the northern part) with a mean of 18\,kpc --- which would also be consistent with the distance estimates given for GCN. The latter objects lie at much farther distances of 30 to 60\,kpc, but they should still be detectable with EBHIS and GASS. Note also that given the geometry and galacto-centric distances, the filament must have penetrated the \ion{H}{i} disk of the MW (at a radius of about 15\,kpc), which of course introduces lots of physical interaction processes that likely are not well described by the simulations.
\begin{figure*}[!t]
\centering
\includegraphics[width=0.8\textwidth,bb=55 47 401 280,clip=]{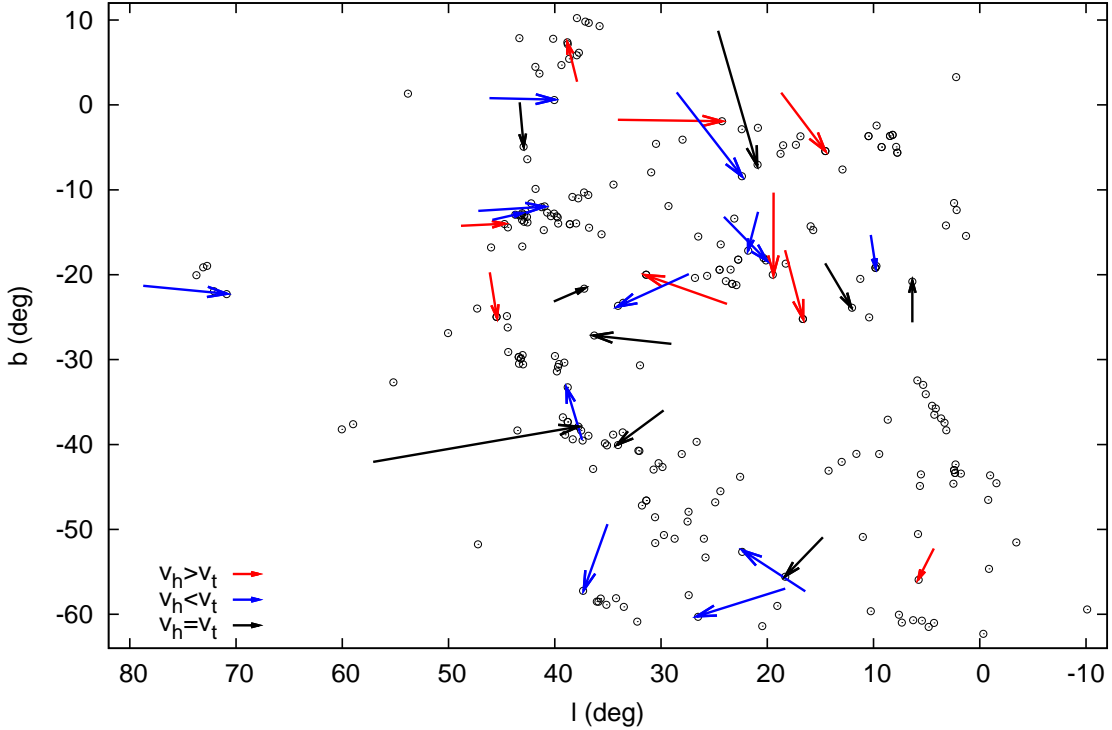}
\caption{Head-tail  vectors as a function of Galactic coordinates. We have differently marked clouds where $v_\mathrm{head}<v_\mathrm{tail}$ (because all clouds move with negative radial velocities, this means that the tail moves slower toward the observer than the head) and $v_\mathrm{head}>v_\mathrm{tail}$. In some cases the velocity fields did not show a clear sign of a separation. The length of each vector indicates the separation of the head and tail but was scaled with a factor of ten for improved visualization.}
\label{figHTvs_lb}%
\end{figure*}

\subsubsection{Direct orbit calculation}
Based on LAB data, \citet{jin10} recently used several clouds at velocities $v_\mathrm{gsr}=-170\ldots-30\,\mathrm{km\,s}^{-1}$ to compute a hypothetical orbit that the gas might follow. This orbit is also shown in Fig.\,\ref{figclumpcoordinatesGN96} marked with the solid line (color again encodes $v_\mathrm{gsr}$). Interestingly, the proposed orbit  describes the spatial position of most of the clouds in the N--S stripe pretty well. However, because the radial velocity scatter is huge along the N--S stripe, the orbit cannot predict the velocities of all features. This could be because the orbit calculation could not take into account the complex physical interaction processes that probably play an important role for GCN (see also Section\,\ref{secdiscussion}).

The orbit predicts a heliocentric distance of the associated GCN clouds in the range of about 15 to 35\,kpc \citep[with farther distances in the south, see also][their figure 4]{jin10}. While the lower limit is consistent with the results from \citet[][]{diaz11}, the upper limit is quite different. Moreover, the sign of the distance gradient is opposite. \citet{jin10} conclude that a dwarf galaxy is likely to be the origin of the tidal stream, though  no optical counterpart was detected along the orbit so far.

\subsection{Head-tail structures as tracer for the motion of clouds}\label{subsecmorphology}

In  Fig.\ref{figHTvs_lb} we show the head-tail vectors as a function of Galactic coordinates. The head of the arrows mark the position of the heads of the clouds.  Most of the HT vectors seem to be randomly distributed, but in the northeast  ($l\sim10\degr\ldots25\degr;~b\sim-30\degr\ldots-5\degr$) we find nine HT vectors with coherent directions that point southward (and one apparently points northward). Furthermore, in the northern part there are six HT vectors pointing westward. Usually the HT vectors are interpreted as motion of the clouds relative to their environment because the more diffuse tail is thought to be created by ram-pressure stripping. In this scenario one would expect the velocity of the tails (relative to the surrounding medium) to be lower than that of the heads. Therefore, we calculated the velocity gradient along the HT vector where possible --- in some cases the superposition of clumps made the velocity fields unusable and in other cases no clear gradient was visible. We have differently marked the vectors in Fig.\ref{figHTvs_lb} according to their velocity gradient. Clouds with $v_\mathrm{head}<v_\mathrm{tail}$ show the expected velocity difference (ram-pressure scenario). Generally, the absolute differences of the head and tail velocities are relatively small (at most $6\,\mathrm{km\,s}^{-1}$).

Interestingly, although the apparent motion of HT clouds in the northeastern and northern regions are coherent, the observed velocity gradients are not consistent with the velocity differences between head and tail expected from  a simple ram-pressure scenario. About half of the vectors have the opposite sign of the velocity gradient. While one could in principle argue that overlapping clumps might mimic head-tail structures, this should lead to a more random alignment of the HT vectors except if there is a preferential alignment of adjacent clumps. This could eventually be explained by gas fragmenting along its motion vector. Then the HT vectors would not mark the motion of the clouds but at least the alignment of the motion. Interferometric measurements of the suspicious clouds might help to solve this puzzle because they would allow for a better distinction of the head and tail. Possibly, the observed effect is even natural. \citet{kalberla06} analyzed the typical radial velocity offsets of cores with respect to their envelope (i.e., cold vs. warm line component) and find that both faster and slower cores have approximately the same frequency of occurrence.

\subsection{Correlations between $\Delta v$, $N^\mathrm{max}_\ion{H}{i}$ and $v_\mathrm{gsr}$}\label{subsec:correlations}
\subsubsection{Full sample}
\begin{figure}
\centering
\includegraphics[width=0.45\textwidth,bb=55 58 395 291,clip=]{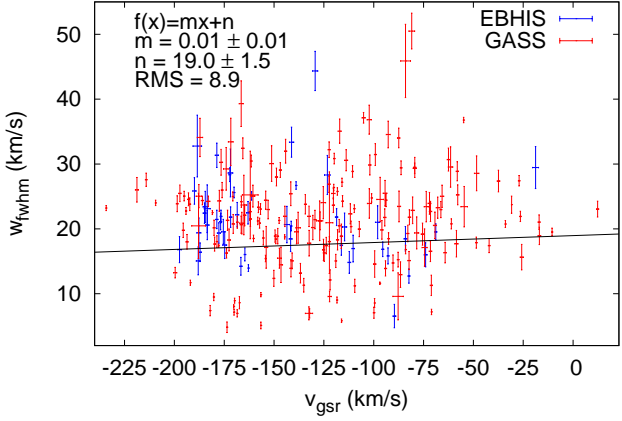}
\caption{Line width of all identified clumps as a function of $v_\mathrm{gsr}$. The solid line is the best-fit linear relation with a value of $0.01\pm0.01\,\mathrm{km\,s}^{-1}\,/\,(\mathrm{km}\,\mathrm{s}^{-1})$.}
\label{fig_w_fwhm_x0_all}%
\end{figure}

In \citet{winkel10b} we reported about a possible line-width -- radial-velocity relation in precursor observations made with EBHIS covering parts of the northern region of complex GCN. Consequently, one of the key questions of this work was to confirm or reject this relation with the help of the EBHIS and GASS data that cover the full GCN field.

Here, using the complete sample of GCN cloudlets, we do not find a significant global $\Delta v$--$v_\mathrm{gsr}$ gradient; see Fig.\,\ref{fig_w_fwhm_x0_all}. 

\begin{figure}
\centering
\includegraphics[width=0.45\textwidth,bb=55 58 395 291,clip=]{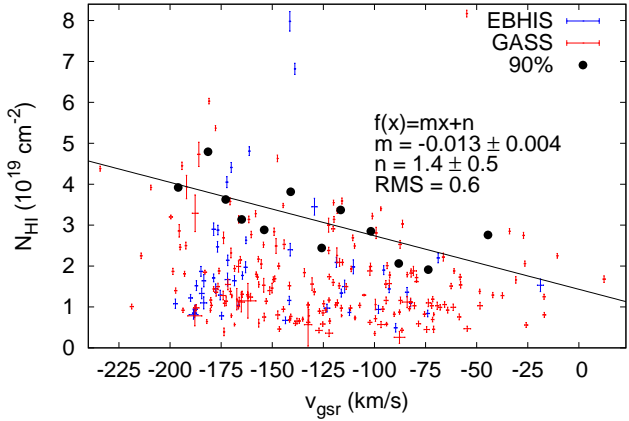}
\caption{Column densities of all identified clumps as a function of $v_\mathrm{gsr}$. To parameterize the upper envelope, the 90\% percentiles for certain velocity intervals were calculated (using variable bin widths such that each bin contains the same number of clouds). The solid line is the best-fit linear relation through these data points, showing a significant gradient of about $-13\pm4\cdot10^{16}\,\mathrm{cm}^{-2}\,/\,(\mathrm{km}\,\mathrm{s}^{-1})$.}
\label{fig_N_HI_x0_all}%
\end{figure}

In Fig.\,\ref{fig_N_HI_x0_all} we show the column densities as a function of radial velocity of the clouds. Apparently, the upper envelope decreases toward lower absolute velocities. To parameterize this behavior, we calculated the 90\% percentiles for certain velocity intervals (using variable bin widths such that each bin contains the same number of clouds). A linear function was fitted to the resulting data points. A significant relation between these upper column densities $N^\mathrm{max}_\ion{H}{i}$ and $v_\mathrm{gsr}$ is obtained of about $-13\pm4\cdot10^{16}\,\mathrm{cm}^{-2}\,/\,(\mathrm{km}\,\mathrm{s}^{-1})$.

However, as can be seen in Figs.\,\ref{figclumpcoordinates} and \ref{figHTvs_lb}, one can hardly speak of a single-cloud population, but there seem to be several distinct regions with respect to $l$, $b$, and $v_\mathrm{gsr}$. 

\subsubsection{Subpopulations}
\begin{figure}
\centering
\includegraphics[width=0.45\textwidth,bb=59 49 396 291,clip=]{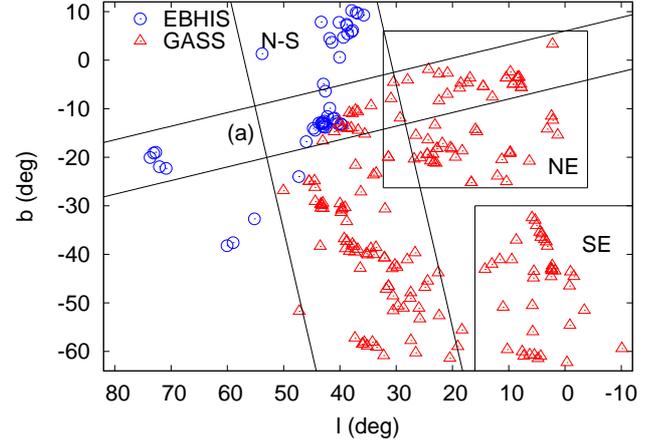}
\caption{For some of our analyses we defined subsets of the cloud catalog. Four regions where used, Stripe N--S, Stripe (a), and the northeastern and southeastern regions. For the motivation of the different regions see text.}
\label{figsplittings}%
\end{figure}

The first very prominent population is located along the filament (hereafter referred to as Stripe N--S) predicted by the GN96+drag model \citep{diaz11}. It manifests itself almost like a ``wall'' in the three-dimensional parameter space because the scatter in $v_\mathrm{gsr}$ is considerable along this curve. Two other regions appear visually distinct, one in northeastern part (denoted as Region NE, having $v_\mathrm{gsr}\lesssim-120\,\mathrm{km\,s}^{-1}$) and one in the southeast (Region SE, $v_\mathrm{gsr}\gtrsim-140\,\mathrm{km\,s}^{-1}$); see Fig.\,\ref{figsplittings}. The northeastern region contains several HT clouds that apparently move southward. Both regions, NE and SE, show much less scatter in radial velocities than clouds in Stripe N--S. Although the mean GSR velocity in the NE and SE regions is different by about $75\,\mathrm{km\,s}^{-1}$ and there is little overlap, the two regions are connected in velocity via the N--S stripe.

We introduce a fourth subpopulation, with objects located along a slice defined by the orientation of the IVC feature (a); compare Fig.\,\ref{figivcfeature}. This slice is hereafter denoted as Stripe (a); see Fig.\,\ref{figsplittings}. Five HT clouds lie within Stripe (a) and surprisingly have their HT vectors closely aligned with the slice direction.

\begin{figure}
\centering
\includegraphics[width=0.45\textwidth,bb=55 80 395 291,clip=]{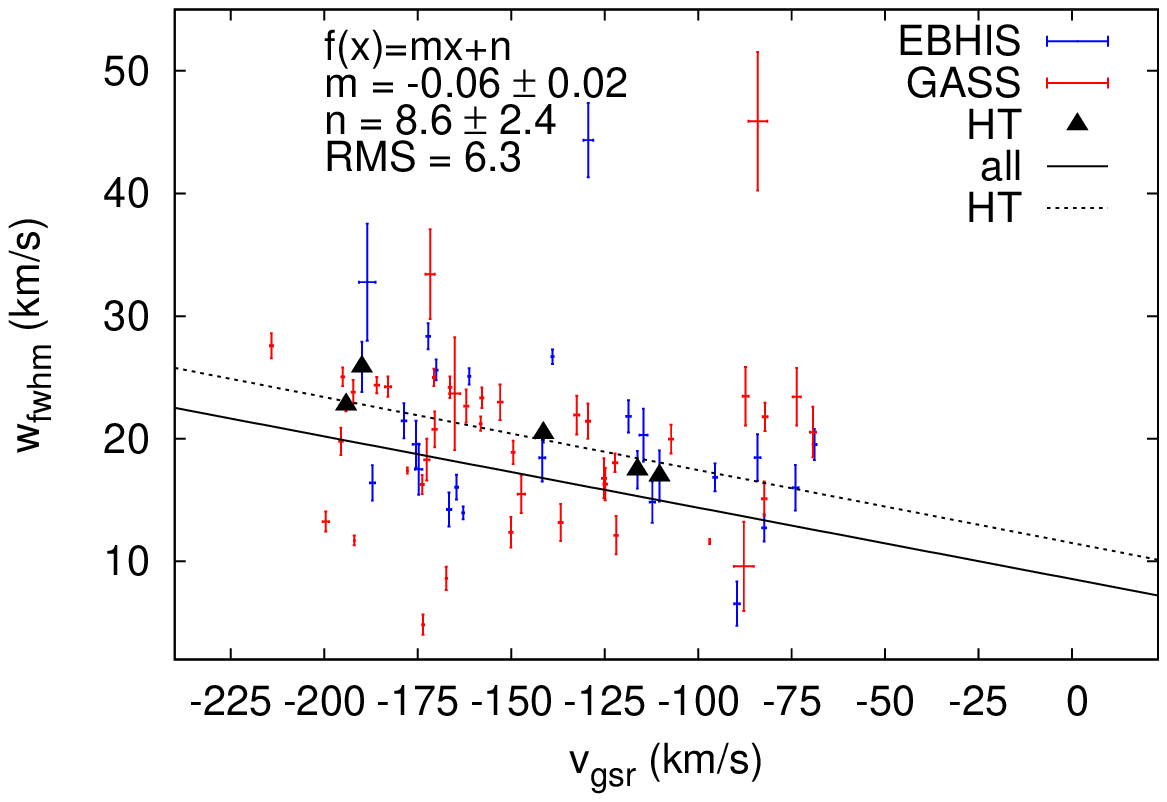}\\[0ex]
\includegraphics[width=0.45\textwidth,bb=55 58 395 291,clip=]{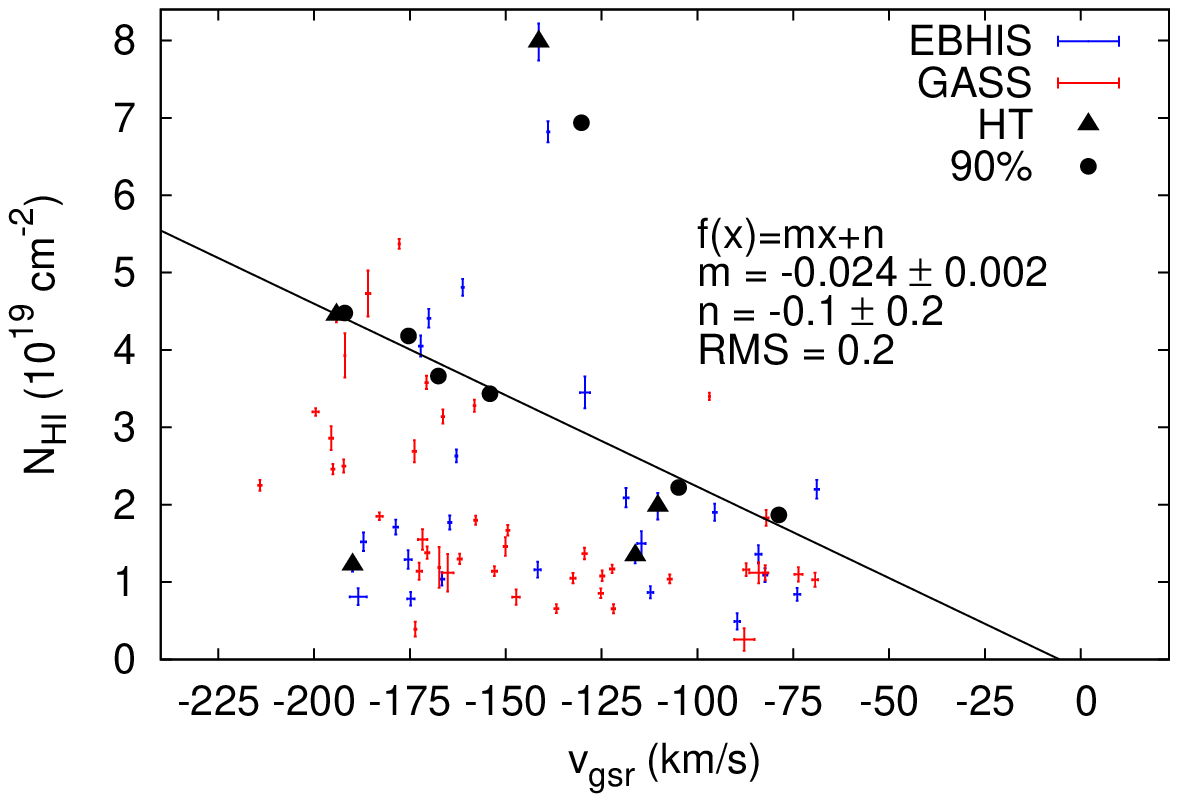}
\caption{Line widths (upper panel) and column densities (lower panel) as a function of radial velocity for clumps along Stripe (a) as defined in Fig.\,\ref{figsplittings}. Head-tail clouds are marked with a triangle. For the line widths we computed a linear fit through all data points (solid line) and through HT clouds only (dotted line), respectively. For the column densities we calculated the 90\% percentiles  as in Fig.\,\ref{fig_N_HI_x0_all} to parameterize the upper envelope. The solid line is the best-fit linear relation of $N^\mathrm{max}_\ion{H}{i}$ vs. $v_\mathrm{gsr}$ neglecting the outlier at $v_\mathrm{gsr}\approx-130\,\mathrm{km\,s}^{-1}$.}
\label{fig_w_fwhm_x0_split1}%
\end{figure}

\begin{table}
\caption{Results of the linear fits for the $\Delta v$--$v_\mathrm{gsr}$ and $N^\mathrm{max}_\ion{H}{i}$--$v_\mathrm{gsr}$ relation for the different (sub)samples. 
The corresponding plots are shown in Fig.\,\ref{fig_w_fwhm_x0_all},\,\ref{fig_N_HI_x0_all}, and \ref{fig_w_fwhm_x0_split1}.}
\label{tabfitresults}
\begin{tabular}{lcccc}\hline\hline
\rule{0ex}{3ex}Sample&$\delta(\Delta v\mathrm{-}v_\mathrm{gsr})$&RMS & $\delta(N^\mathrm{max}_\ion{H}{i}\mathrm{-}v_\mathrm{gsr})$ &RMS \\ 
\rule{0ex}{4ex}&$\displaystyle\frac{\mathrm{km\,s}^{-1}}{\mathrm{km}\,\mathrm{s}^{-1}}$&$\mathrm{km\,s}^{-1}$    &$\displaystyle\frac{\mathrm{cm}^{-2}}{\mathrm{km}\,\mathrm{s}^{-1}}$ & $\mathrm{cm}^{-2}$\\[2ex]\hline
\rule{0ex}{3ex}Complete&$1\pm1\cdot10^{-2}$&$9$&$-13\pm4\cdot10^{16}$&$6\cdot10^{18}$\\
Stripe N--S&$2\pm1\cdot10^{-2}$&$8$&$-10\pm2\cdot10^{16}$&$2\cdot10^{18}$ \\
Stripe (a)&$-6\pm2\cdot10^{-2}$&$6$&$-24\pm2\cdot10^{16,\dagger}$&$2\cdot10^{18,\dagger}$ \\
HT (a)&$-6\pm1\cdot10^{-2}$&$1$&& \\
Region NE&$-8\pm4\cdot10^{-2}$&$9$&$-20\pm20\cdot10^{16}$&$9\cdot10^{18}$ \\
Region SE&$-5\pm8\cdot10^{-2}$&$10$&$-12\pm11\cdot10^{16}$&$5\cdot10^{18}$ \\
\hline\\[0ex]
\end{tabular}
$^\dagger$ Neglecting an outlier, see Fig.\,\ref{fig_w_fwhm_x0_split1}.
\end{table}

Again, we parameterized possible $\Delta v$--$v_\mathrm{gsr}$ and $N^\mathrm{max}_\ion{H}{i}$--$v_\mathrm{gsr}$ correlations using linear regression. Table\,\ref{tabfitresults} shows a summary of all relevant fit parameters for the different subsamples. 

\begin{figure}
\centering
\includegraphics[width=0.45\textwidth,bb=57 50 395 291,clip=]{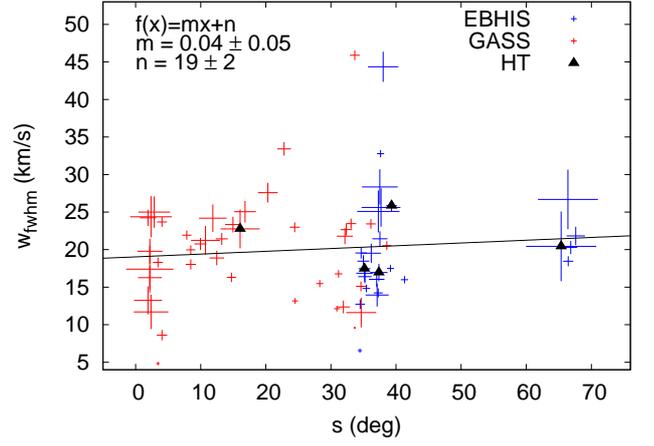}
\caption{Line widths as a function of spatial position (measured from the east) for clumps along Stripe (a) as defined in Fig.\,\ref{figsplittings}.}
\label{fig_w_fwhm_dist_split1new}%
\end{figure}

Except for Stripe (a) we find no significant correlation of $\Delta v$ vs. $v_\mathrm{gsr}$. Stripe (a) also shows high evidence for a column density evolution with  $v_\mathrm{gsr}$. Because Stripe (a) contains several HT clouds that are morphologically aligned with the CND feature (a), we additionally had a look at the column densities and line widths of the HT clouds (marked with triangles in Fig.\,\ref{fig_w_fwhm_x0_split1}). In the line width--velocity panel the HT clouds are remarkably tight-correlated but we emphasize that only five data points are available. Hence, we performed a Monte-Carlo test where we randomly drew five data points out of the Stripe (a) sample. In only 3\% of all cases we obtained a correlation with similar quality of fit (in terms of reduced $\chi^2$ value), which makes it unlikely that the observed relation in the HT subsample of Stripe (a) is just chance. We also point out that the correlation of the line width with the radial velocity is not related to a possible evolution of the line widths with spatial position (although there is some evolution of $v_\mathrm{gsr}$ with position, as seen in Fig.\,\ref{figclumpcoordinates}). To show this, we plot in Fig.\,\ref{fig_w_fwhm_dist_split1new} the line widths as a function of position along Stripe (a), which do not show a  correlation. Apparently, the line width--velocity relation is a genuine property independent of the spatial position of the clouds.

The question remains why all other subsamples do not show a $\Delta v$--$v_\mathrm{gsr}$ correlation. One could argue that the complexity of the whole field makes it difficult to find the correct subsets of clouds that are  physically connected. This is at least suggested from Fig.\,\ref{figHTvs_lb}, where it appears that apart from few regions the dynamical motions are quite stochastic. 

Although no reliable $\Delta v$--$v_\mathrm{gsr}$ relation could be found for all regions except Stripe (a), the maximal column densities vary significantly with $v_\mathrm{gsr}$. Several of the regions (Stripe (a) and N--S as well as the complete sample) show a significant decrease for the upper envelope of the column densities with lower absolute values of the radial velocities. This might be a sign of ongoing interaction with the ambient halo material that dissolves and decelerates the clouds.

\subsection{Statistical properties of isolated clouds --- substructure}\label{subsecsmallscalestructure}
In this subsection we discuss the statistical properties of clumps that are isolated, i.e., not part of a filament or overlapping with other clouds. This allows us to introduce another cloud parameter, $\alpha_\mathrm{sub}$, which shall be an indicator for the amount of substructure for a clump. If a cloud were point-like (or had a Gaussian-like column density profile), it would produce a Gaussian-shaped column density profile in the data as the telescopes performs a convolution of the true brightness distribution with the telescope beam. Fitting a Gaussian to the data should then lead to a residual that exhibits the same noise level as the surrounding (emission-free) regions. On the other hand, when substructure is present, a Gaussian would only provide a poor fit to the column densities, hence the residual noise level should be higher than before. We define 
\begin{equation}
\alpha_\mathrm{sub}\equiv \frac{\sigma_\mathrm{rms}(\mathrm{fit~residual})}{\sigma_\mathrm{rms}(\mathrm{surrounding})}
\end{equation}
as the residual noise level after fitting 2D Gaussians to the clumps relative to the noise of emission free regions. Unfortunately, owing to the limited angular resolution of the single-dish data, one can treat the amount of substructure only in this statistical way.
\begin{figure}
\centering
\includegraphics[width=0.45\textwidth,bb=3 15 393 408,clip=]{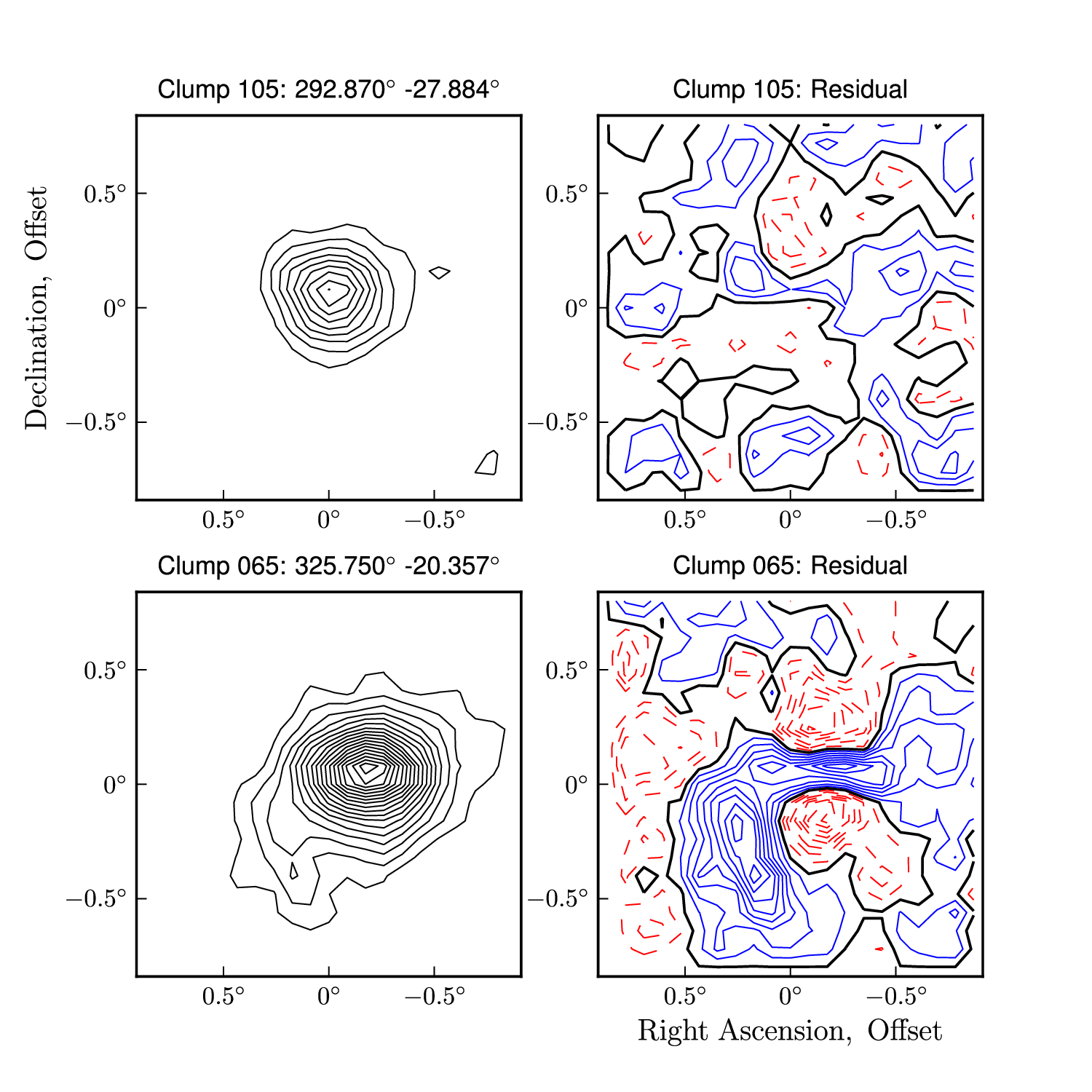}
\caption{For two isolated clumps a two-dimensional Gaussian was fitted to the column density profile (left panels). The residual (right panels) shows different behavior, for the upper row the noise level is very similar to an emission-free region in the surrounding, the lower row exhibits remaining features increasing the measured noise value. The residual RMS can be used as an estimator to the amount of substructure within a clump. Contour lines in the left panels start at $3\sigma_\mathrm{rms}$ and have a spacing of $3\sigma_\mathrm{rms}$. Contour lines in the right panels are spaced with $1\sigma_\mathrm{rms}$, solid contours mark positive values, dashed contours are negative, the thick solid lines are zero flux.}
\label{figclumpstatsexamples}%
\end{figure}

In Fig.\,\ref{figclumpstatsexamples} we show two examples: a clump that is fairly well-fitted by a single Gaussian (top panels) and another clump that has enhanced noise in the residual (bottom panels). In the former case we obtain $\alpha_\mathrm{sub}=1.1$, while for the latter $\alpha_\mathrm{sub}=2.5$. Most of the isolated clouds in our catalog are located in the southern hemisphere. Therefore, we restricted the following analysis to the GASS data, because it is not clear in which way the different beam sizes and noise levels of EBHIS and GASS would introduce systematic effects. In total the subset of isolated clumps contains 63 objects. 

\begin{figure}
\centering
\includegraphics[width=0.45\textwidth,clip=]{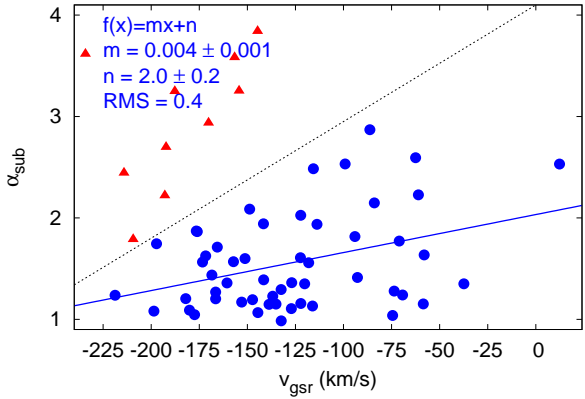}
\caption{Substructure estimator $\alpha_\mathrm{sub}$ for all isolated clumps as a function of $v_\mathrm{gsr}$. There appear to be two independent populations, which we separated using the dotted line. For the circles a linear function was fitted to the data, showing a gradient of $4\pm1\cdot10^{-3}\,(\mathrm{km\,s}^{-1})^{-1}$.}
\label{figclumpvgsr_substructure}%
\end{figure}

\begin{figure}
\centering
\includegraphics[width=0.45\textwidth,bb=66 61 426 262,clip=]{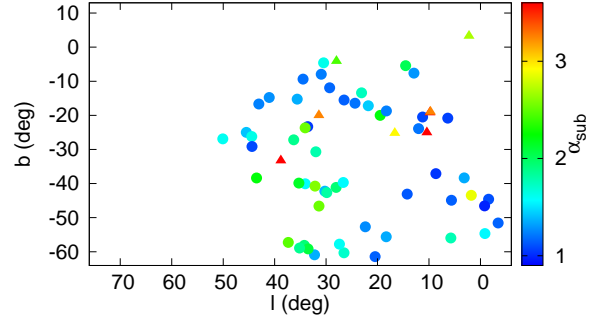}
\caption{Positions of isolated clumps in Galactic coordinates. Triangles and circles mark the two populations defined in Fig.\,\ref{figclumpvgsr_substructure}. The substructure estimate $\alpha_\mathrm{sub}$ is color-coded. For the population marked with circles we find a gradient of $\alpha_\mathrm{sub}$ in N--S direction with higher values in the southern part of complex GCN.}
\label{figclumpcoordinatessubstructure}%
\end{figure}

In Fig.\,\ref{figclumpvgsr_substructure} we show $\alpha_\mathrm{sub}$ as a function of $v_\mathrm{gsr}$. There seem to be two different populations, which we separated for some of the following analyses. For clumps below the dotted line (blue circles) a correlation of $\alpha_\mathrm{sub}$ and $v_\mathrm{gsr}$ is found: $\delta(\alpha_\mathrm{sub}\mathrm{-}v_\mathrm{gsr})=4\pm1\cdot10^{-3}\,(\mathrm{km\,s}^{-1})^{-1}$. In Fig.\,\ref{figclumpcoordinatessubstructure} $\alpha_\mathrm{sub}$ is plotted vs. Galactic coordinates. Triangles and circles again mark the different populations as defined in Fig.\,\ref{figclumpvgsr_substructure}. There appears to be no clear spatial separation between both subsets. For one of the subsets (circles) one can see a trend in N--S direction of $\alpha_\mathrm{sub}$ increasing toward the south. While one could argue that our method of defining the amount of substructure might be subject to systematic effects and uncertainties, it is remarkable that this general trend is observed, which can hardly be explained by coincidence.

\begin{figure}
\centering
\includegraphics[width=0.45\textwidth,bb=66 61 426 262,clip=]{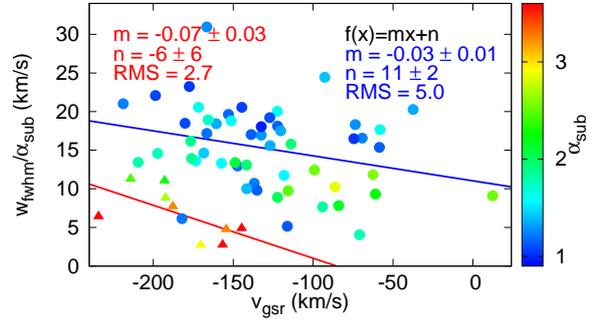}
\caption{Quantity $\frac{\Delta v_\mathrm{FWHM}}{\alpha_\mathrm{sub}}$ as a function of $v_\mathrm{gsr}$ for isolated clumps. Triangles and circles mark the two populations defined in Fig.\,\ref{figclumpvgsr_substructure}. The substructure estimate $\alpha_\mathrm{sub}$ is color-coded. For both populations linear functions were fitted to the data points. A bifurcation is observed.}
\label{figclumpx0gsr_fwhmdivsubstr}%
\end{figure}

There is no appreciable trend of $\alpha_\mathrm{sub}$ with the line width of the clouds but the correlation of $\alpha_\mathrm{sub}$ and $N_\ion{H}{i}$ although the scatter is large (not shown here). For comparison we also computed the line width--radial velocity relation for the subsample of isolated clouds, which does not exhibit any clear correlation. However, based on the idea that the amount of substructure might increase the effectively measured line width, we display in Fig.\,\ref{figclumpx0gsr_fwhmdivsubstr} the quantity $\frac{\Delta v_\mathrm{FWHM}}{\alpha_\mathrm{sub}}$ as a function of $v_\mathrm{gsr}$ and observe, first, a bifurcation of both populations (circles and triangles) and, second, a correlation  of this "effective" line width with radial velocity, which is less scattered about the linear fit than before, even so we did not divide the catalog into different subsamples based on their spatial position.

Note that the observed relation between $\alpha_\mathrm{sub}$ and $N_\ion{H}{i}$ can be explained in a natural way because two similarly distorted objects with different amplitudes would not necessarily lead to similar values of $\alpha_\mathrm{sub}$ because the residual scatter is expected to be higher if the Gaussian has a larger amplitude. However, in that case similar results as shown in Fig.\,\ref{figclumpx0gsr_fwhmdivsubstr} should be obtained when using  $\frac{\Delta v_\mathrm{FWHM}}{N_\ion{H}{i}}$ instead of $\frac{\Delta v_\mathrm{FWHM}}{\alpha_\mathrm{sub}}$, but this is not the case. Also, the two different populations are not as clearly separated in the $N_\ion{H}{i}$--$v_\mathrm{gsr}$ space as in $\alpha_\mathrm{sub}$--$v_\mathrm{gsr}$. It seems that the substructure estimator provides a more genuine property than the column density despite their apparent correlation.

\section{Discussion}\label{secdiscussion}
The observed properties of clouds in our sample show that GCN is quite different compared to other large HVC complexes. Not only the small spatial size of most of the clumps and the low fraction of cold-line components are rather untypical, but also the missing diffuse extended \ion{H}{i} emission, which is detected in most other HVC complexes \citep[e.g.,][using the less sensitive LAB survey data]{kalberla06}. This is especially true because the absorption line data indicate the presence of a (low-column density) diffuse gas phase.  Furthermore, we find many signs for collisionally or photo-ionized gas in the region. Several head--tail features were observed, and we find a decrease of the maximal \ion{H}{i} column densities toward lower absolute radial velocities. This makes it likely that GCN clouds are subject to interaction processes with an ambient medium (i.e., the Milky Way halo). 

Unfortunately, the large distance uncertainty does not allow us to estimate important physical parameters like temperature or pressure with sufficient accuracy to rule out or accept the various possible interaction processes directly, i.e. Rayleigh--Taylor- and Kelvin--Helmholtz instabilities, ram-pressure, or shocks (which could also lead to fragmentation). However, we will discuss below some of the mechanisms under the assumption that clouds in the N--S stripe are indeed related to the Leading Arm, as suggested by the simulations of \citet{diaz11}, which constrains the distance to 14 to 22\,kpc. This is at least consistent with earlier (rough) distance estimates and partially with the orbit proposed by \citet{jin10}. 

In this case, for a beam-sized (unresolved) cloudlet at 18\,kpc distance with a typical column density of $10^{19}\,\mathrm{cm}^{-2}$ we estimate the volume density to be on the order of $n_\mathrm{c}=0.1\,\mathrm{cm}^{-3}$. The size of such an object would be about 50\,pc. Furthermore, using the MW model of \citet{kalberla07}, we can obtain an estimate of the pressure and density of the surrounding halo gas along the N--S filament: $P/k_\mathrm{B}\sim10^3\,\mathrm{K\,cm}^{-3}$ and $n_\mathrm{h}\gtrsim10^{-2}\,\mathrm{cm}^{-3}$ close to the MW disk and $P/k_\mathrm{B}\lesssim10^2\ldots10^3\,\mathrm{K\,cm}^{-3}$ and $n_\mathrm{h}\sim10^{-4}\ldots10^{-3}\,\mathrm{cm}^{-3}$ at higher Galactic latitudes. Typical masses of the larger structures, e.g., at $(l,b)=(37\degr,-38\degr)$, would be few times $10^4\,M_\odot$, the smaller clouds could have a few $10^3\,M_\odot$.

In their analysis of the clouds at the tip of the MS, \citet{stanimirovic08} estimate time and length scales for the physical processes important for interacting clouds. While at a first glance our objects have similar properties (i.e., density, size, and line widths), the surrounding medium makes a big difference. Particularly, we observe signs of ongoing ram-pressure interaction (the head--tail structures), which were absent in the MS sample. This is not surprising, given the high densities of the halo material at smaller distances compared to the MS. Hence, in our case the Kelvin--Helmholtz time scales are larger (about 2\,Gyr), while thermal instabilities in principle should have similar time scales of a few tens of Myr. 

\citet{heitsch09} used simulations to study the stability of neutral clouds against ram-pressure for different scenarios (wind-tunnel and free-fall in the MW potential). While both cases are not exactly applicable for the GCN clouds, some of the authors' basic conclusions might be of interest for us. First, if the initial masses are on the order of $10^3$ to $10^4\,M_\odot$, the time scale for complete disruption is on the order of 50 to at most 100\,Myr (cloud properties in our sample are similar to their simulations and the environmental densities are likely even higher). Although cooling times (and hence the time scale for thermal instabilities) are likely shorter or  comparable, the wind-tunnel simulations show no significant cooling of the clouds. Because the clouds loose material through the ram-pressure interaction, cooling is even less effective. This could explain the small fraction of cold objects in our sample. Note that the ablated material might perhaps form smaller fragments, but these would have much smaller masses than the main cloud. A second important point is that the fragments are most likely not stable (because of the less efficient cooling) and will eventually evaporate. We conclude that ram-pressure interaction can most likely explain the observed cloud properties, though simulations better suited to the GCN scenario would be appreciated.

Under these considerations, complex GCN might be a prime example for warm gas accretion onto the Milky Way, where neutral \ion{H}{i} clouds or a stream of gas are encountering the Milky Way gas halo. Under certain physical conditions clouds are no longer stable against interaction with the ambient medium, might dissolve and become ionized prior to accretion. This scenario is also suggested by recent simulations \citep[e.g.,][]{blandhawthorn09}. In that case the GCN objects would just be in a transient state. This would match our observed  $N^\mathrm{max}_\ion{H}{i}$--$v_\mathrm{gsr}$ relation. The lack of \ion{Na}{i}, which is a tracer for the colder cores of typical clouds, is another sign that the typical life times of the clouds might not be long enough to allow for substantial cooling of the cores --- which explains the under-abundance of cold clouds in our sample compared to most other HVC complexes.

If the hypothesis is true, clouds would not be visible anymore, which immediately raises the question of how much ionized gas mass is contained in complex GCN. Because it is hard to observe, it is still unclear today whether the (global) ionized fraction of accreted material is able to explain the observed constant star-formation rate of the MW. Metal absorption line studies \citep[using large samples; see][]{benbekhti08,shull11} of the HVC sky will hopefully shed light on this issue. One should also keep in mind that \ion{Ca}{ii} was detected in sight lines without \ion{H}{i} emission, hence, there definitely is low-column density material.

We do not only observe a general relation between the upper envelope of the column densities and the radial velocity, but this relation is also significantly found for two subregions (Stripe (a) and N--S). Although the other two subsamples (Region NE and SE) do not individually show a significant relation, all of the slopes are negative, which increases the overall significance. The correlation for Stripe (a) and N--S is surprisingly tight considering the huge scatter in the observed radial velocities in these regions, while for the regions NE and SE, which have much less scatter in radial velocities, the relation is much less constrainted. Furthermore, given the observed (non-)alignments of the HT vectors in each of the regions, we conclude that clouds in the four subsamples may have different origins.

One striking result is the detection of five HT clouds in Stripe~(a) --- the definition of which was based only on the IVC feature near the CND --- pointing eastward. This is compatible with the observed $N^\mathrm{max}_\ion{H}{i}$--$v_\mathrm{gsr}$ relation under the hypothesis that clouds with lower velocities, i.e., which were decelerated for a longer period of time, have already lost more gas through ram pressure. Here, we cannot prove a causal relationship, but it could be that the IVC feature is just the oldest component within Stripe~(a), which has just encountered the MW disk. 

\section{Summary and outlook}\label{secsummary}
The analysis has shown that complex GCN is indeed a very complex object consisting of more than two hundred (mostly individual) tiny clouds, which are often unresolved even with the much better angular resolution of the EBHIS and GASS data compared to any previous study of GCN. We do not observe  diffuse extended 21-cm emission around the isolated clouds, although the column density detection limit of our data would be sufficient to detect this gas phase at reasonable distances. Our observations suggest that the clouds are likely a transient phenomenon in a very turbulent and highly dynamic environment as expected if GCN is really the result of ongoing accretion. It might be that the clouds become ionized prior to accretion, which might render a large fraction of the total gas mass in GCN invisible for the \ion{H}{i} survey data. The fact that we detect \ion{Ca}{ii} signatures and other ionized species at velocities consistent with the \ion{H}{i} structures supports the hypothesis that there is a significant amount of low-column density material around.

The presented work is only the first step toward a deeper understanding of complex GCN as a whole. We are going to use interferometric observations for studying the morphologically interesting cases  at much better angular resolution. This would allow us to obtain better constraints on the true motion of these clouds, because currently there is often an ambiguity between the morphology and observed radial velocity difference between head and tail. It would also be interesting whether unresolved clumps further fragment into individual cores, because this could provide a consistency check for our substructure estimator. Furthermore, deep single-dish observations of some smaller regions could shed light on the question of the missing diffuse extended component. If it turns out that the latter is not physically present, it might be worthwhile to search for additional absorption lines to study the  ionized gas phase to estimate its total mass contribution.

While this work mostly dealt with the general (statistical) properties of our cloud sample, we will next analyze potential subpopulations in more detail (Darmst\"{a}dter in prep.). Based on our findings, we suspect that GCN is a mixture of various objects with different origins, which one would need to separate to understand the complex as a whole.

\begin{acknowledgements}
The authors thank the Deutsche Forschungsgemeinschaft (DFG) for financial support under the research grant KE757/7-1 and KE757/9-1. Our results are based on observations with the 100-m telescope of the MPIfR (Max-Planck-Institut f\"ur Radioastronomie) at Effelsberg. We would like to thank J. Diaz and K. Bekki for kindly providing us access to their simulation data of the Magellanic System, and the referee, B. Wakker, for his very useful comments and suggestions, which helped us to improve the manuscript.
\end{acknowledgements}

\bibliographystyle{aa}
\bibliography{references}

\end{document}